\documentclass[%
reprint,
superscriptaddress,
amsmath,amssymb,amsthm,mathrsfs,
aps,
prl,
]{revtex4-2}

\usepackage{graphicx}
\usepackage{dcolumn}
\usepackage{bm}
\usepackage[usenames]{color}
\usepackage{soul}
\usepackage[ 
margin=0.75in,
]{geometry}
\usepackage[normalem]{ulem}

\graphicspath{{./Figures/}}

\def\muSR{$\mu$SR}

\begin{document}
	
	\title{
		Tuning the spin dynamics and magnetic phase transitions of the Cantor alloy via composition and sample processing protocols: A muon spin relaxation study
	}
	
	\author{Emma Zappala}
	\affiliation{ %
		Department of Physics and Astronomy, Brigham Young University, Provo, Utah 84602, USA.}
	
	\author{Timothy A.\ Elmslie}
	\affiliation{Department of Physics, University of Florida, Gainesville, FL 32611-8440, USA}
	
	\author{Gerald D. Morris}
	\affiliation{Centre for Molecular and Materials Science, TRIUMF, Vancouver, British Columbia, Canada V6T 2A3}
	
	\author{\\Mark~W.\ Meisel}
	\affiliation{Department of Physics, University of Florida, Gainesville, FL 32611-8440, USA}
	\affiliation{National High Magnetic Field Laboratory, University of Florida, Gainesville, Florida 32611-8440, USA}
	
	\author{R\'emi Dingreville}
	\affiliation{Center for Integrated Nanotechnologies, Sandia National Laboratories, Albuquerque, NM USA}

	\author{James J.\ Hamlin}
	\affiliation{Department of Physics, University of Florida, Gainesville, FL 32611-8440, USA}
	
	\author{Benjamin A. Frandsen}
	\affiliation{ %
		Department of Physics and Astronomy, Brigham Young University, Provo, Utah 84602, USA.}
	\email{benfrandsen@byu.edu}

	\begin{abstract}
		CrMnFeCoNi, also called the Cantor alloy, is a well-known high-entropy alloy whose magnetic properties have recently become a focus of attention. We present a detailed muon spin relaxation study of the influence of chemical composition and sample processing protocols on the magnetic phase transitions and spin dynamics of several different Cantor alloy samples. Specific samples studied include a pristine equiatomic sample, samples with deficient and excess Mn content, and equiatomic samples magnetized in a field of 9~T or plastically deformed in pressures up to 0.5~GPa. The results confirm the sensitive dependence of the transition temperature on composition and demonstrate that post-synthesis pressure treatments cause the transition to become significantly less homogeneous throughout the sample volume. In addition, we observe critical spin dynamics in the vicinity of the transition in all samples, reminiscent of canonical spin glasses and magnetic materials with ideal continuous phase transitions. Application of an external magnetic field suppresses the critical dynamics in the Mn-deficient sample, while the equiatomic and Mn-rich samples show more robust critical dynamics. The spin-flip thermal activation energy in the paramagnetic phase increases with Mn content, ranging from $3.1(3) \times 10^{-21}$~J for 0\% Mn to $1.2(2) \times 10^{-20}$~J for 30\% Mn content. These results shed light on critical magnetic behavior in environments of extreme chemical disorder and demonstrate the tunability of spin dynamics in the Cantor alloy via chemical composition and sample processing.
	\end{abstract}
	
	\maketitle
	
	\section{Introduction}	
	
	High-entropy alloys (HEAs), typically defined as alloys made up of five or more elements in equal or near-equal proportions, have driven intense research efforts for roughly two decades now~\cite{Yeh2004, tsai2014high, georg;nrm19}. They possess unique and often advantageous mechanical and functional properties compared to conventional alloys, such as high specific strength, extreme fracture toughness and ductility, and corrosion resistance~\cite{ ye2016high,li;pms21}. Furthermore, the compositional space of HEAs is vast, offering practically limitless possibilities for discovering new materials with exceptional properties. Recently, magnetic properties of HEAs have also attracted interest, both for fundamental studies of magnetism in environments of extreme disorder as well as possibly novel applications enabled by highly tunable magnetic properties in HEAs~\cite{kumar;jmmm22}.
	
	One of the most widely studied HEAs is CrMnFeCoNi, also known as the Cantor alloy~\cite{cantor2004microstructural, canto;pms21}. It has a face-centered cubic (fcc) structure~\cite{zhang2019microstructure} and exhibits excellent ductility \cite{salishchev2014effect} and high fracture toughness \cite{gludovatz2016exceptional}. However, the magnetic properties of the Cantor alloy have become a research focus only in the last few years, and the scientific community's understanding of magnetism in this system is far from complete. Proposed magnetic phases in the Cantor alloy include an antiferromagnetic or spin-glass transition below 25~K\cite{jin2016tailoring}, a spin-glass transition at 93~K and ferromagnetic transition at 38~K~\cite{schneeweiss2017magnetic}, or ferrimagnetic order below 85~K and cluster-glass behavior below 43.5~K~\cite{kamarad2019effect}, illustrating the lack of consensus in the field and raising questions about sample quality and/or reproducibility.
	
	A more coherent picture has begun to emerge based on recent systematic studies~\cite{elmsl;prb22, elmsl;prb23} of numerous compositional variants of the Cantor alloy with multiple experimental and theoretical techniques, including magnetometry, compositional and structural characterization, Hall effect, specific heat, density functional theory (DFT), and muon spin relaxation/rotation (\muSR). For the equiatomic Cantor alloy, magnetic transitions at 85~K and 43~K have been confirmed to be intrinsic to the material, with the upper transition being spin-glass-like and the lower transition best explained as a weak ferrimagnetic transition. A large and temperature-independent contribution to the magnetic susceptibility is attributed to a strong Stoner enhancement to the paramagnetism. Notably, the magnetic properties show a pronounced dependence on the sample processing history, with significant differences observed between as-cast samples, samples that were cold-worked in a hydraulic press, and samples that were annealed at high temperature~\cite{elmsl;prb22}. This history sensitivity helps explain the disparate experimental results reported previously in the literature.
	
	Additionally, systematic control over the magnetic properties has been demonstrated by adjusting the stoichiometry away from the equiatomic composition~\cite{elmsl;prb23}. As the concentration of species that are antiferromagnetic in elemental form (i.e. Mn and Cr) increases, the net effective moment decreases and the spin-glass-like transition shifts to higher temperatures. In the other direction, lower concentrations of Mn and Cr relative to the intrinsically ferromagnetic species Fe, Ni, and Co result in an increase in the net effective moment, and the lower ferrimagnetic transition becomes ferromagnetic. The spin-glass-like transition temperature can be tuned between 55~K and 190~K by modifying the composition, while the lower transition remains stable around 43~K regardless of composition.
	
	These systematic studies have demonstrated predictable and controllable magnetic behavior that can be rationalized on the basis of the chemical composition of Cantor alloys. However, many fundamental aspects of the magnetism in these materials have not yet been explored, including how the underlying spin dynamics and the homogeneity of the magnetic transitions are affected by composition and sample processing history. Understanding these crucial details often requires the use of specialized experimental techniques, such as \muSR~\cite{hilli;nrmp22}. As a local magnetic probe, \muSR\ is sensitive to both long- and short-range magnetic correlations and can reliably quantify the volume fraction of local regions in the sample exhibiting static magnetism versus dynamically fluctuating and/or no magnetism, making it an ideal tool for investigating magnetism in complex materials like HEAs. Furthermore, \muSR\ probes spin dynamics on time scales faster than magnetometry techniques and slower than neutron scattering, making it complementary to these workhorse techniques.
	
	In this article, we build upon our previous work in Refs.~\onlinecite{elmsl;prb22,elmsl;prb23} with more extensive \muSR\ experimentation on multiple Cantor alloy samples spanning a range of compositions and sample processing protocols. More specifically, in addition to a pristine sample of the equiatomic Cantor alloy, other equiatomic samples were subjected to plastic deformation via cold working or pre-magnetized with a large magnetic field. The study also includes samples with excess or deficient Mn concentrations. The \muSR\ results confirm the strong dependence of the magnetic transitions on composition and reveal new information about the effect of composition and sample processing on the magnetic homogeneity and spin dynamics of Cantor alloys. In particular, cold working the equiatomic Cantor alloy results in a significantly more inhomogeneous magnetic transition that occurs gradually throughout the sample volume over an extended temperature window, and deviating from the equiatomic composition suppresses the critical spin dynamics in the vicinity of the transition. Finally, the activation energy of spin-flip processes increases with the antiferromagnetic character of the sample, i.e. with increasing Mn concentration. These findings deepen our understanding of magnetism in Cantor alloys and further highlight the tunability of the magnetic properties via composition and sample processing. 
	
	
	\section{Experimental Details}
	\subsection{Sample synthesis, nomenclature, and bulk characterization}
	The samples studied in this work represent a subset of samples investigated in Ref.~\onlinecite{elmsl;prb22, elmsl;prb23}, where full details regarding the synthesis and characterization can be found. Briefly, elemental Cr, Mn, Fe, Co, and Ni were melted together in the desired stoichiometric ratio in an Edmund Buhler MAM-1 compact arc melter. Samples were melted five times, flipping them over between each melt to promote sample homogeneity. Samples were then sealed in quartz tubes in an Ar atmosphere and annealed at 1100~$^{\circ}$C for six days, after which they were quenched in water. 
	
	Results from five samples are presented in this work. Three samples are from the same growth of equiatomic CrMnFeCoNi known as ``anneal A'' in Ref.~\onlinecite{elmsl;prb22}. Of these three samples, one was flattened in a hydraulic press three times with a pressure of about 0.5~GPa, which we call the ``cold-worked'' sample; another sample was magnetized at 9~T at 300~K before any \muSR\ measurements, which we call the ``premag'' sample; and the third was not processed any further between the annealing and the \muSR\ experiments, which we call the ``pristine'' sample. Two non-equiatomic samples were synthesized, which we label by their Mn content: the sample designated ``Mn\(_{30}\)'' has the composition Mn\(_{30}\)(CrFeCoNi)\(_{70}\), while ``Mn\(_0\)'' has the composition CrFeCoNi. These two samples are included in the studies presented in Ref.~\onlinecite{elmsl;prb23}.
	
	The samples were found to be fcc and single-phase by laboratory x-ray diffraction, as reported in Refs.~\onlinecite{elmsl;prb22,elmsl;prb23} (data not shown here). The magnetic susceptibility was measured via SQUID magnetometry using a Quantum Design Magnetic Properties Measurement System, also first reported in Refs.~\onlinecite{elmsl;prb22,elmsl;prb23}. In this work, we show relevant susceptibility data, some of which were published in these two previous papers.
	
	\subsection{Muon spin relaxation/rotation}
	
	In a \muSR\ experiment, spin-polarized positive muons are implanted in the sample, where they typically come to rest at an interstitial position that minimizes the energy. The muon spin precesses in the local magnetic field at the muon position until it decays into a positron and two neutrinos after a mean lifetime of 2.2~$\mu$s. The positron is emitted preferentially in the direction in which the muon spin was pointing at the time of decay. As such, tracking positron emission directions as a function of time after implantation for a large number of muons provides information about the time evolution of the muon ensemble spin polarization, which in turn yields insight into the local magnetic field distribution at the muon stopping positions. More specifically, pairs of positron detectors are placed on opposite sides of the sample to determine the time-dependent \muSR\ asymmetry $a(t)$, given as 
	\begin{equation}
		\label{eq:asy}
		a(t) = \frac{N_1(t) - N_2(t)}{N_1(t)+N_2(t)},
	\end{equation}
	where $N_1(t)$ and $N_2(t)$ are the number of positron events, corrected for random background counts, recorded in the two detectors at a time $t$ after muon implantation. This quantity is proportional to the muon ensemble spin polarization vector projected onto the axis defined by the detector positions. 
	
	The \muSR\ experiments reported in this study were conducted at the TRIUMF Laboratory in Vancouver, Canada. We used the LAMPF spectrometer on the M20D beamline. The samples were mounted on a low-background copper sample holder and the temperature was controlled using a helium gas flow cryostat. Data were collected in zero-field (ZF) mode, in which no external magnetic field was applied and background fields are typically below $2\times10^{-6}$~T; in longitudinal-field (LF) mode, in a which a magnetic field up to 0.4~T was applied parallel to the initial muon spin direction; and weak-transverse-field (wTF) mode, in which a weak field of 0.003~T was applied perpendicular to the initial muon spin direction. The total number of positron events recorded by the two detectors during each run was approximately 10 million for ZF and LF and 3 million for wTF. We analyzed the \muSR\ data using the open-source software BEAMS~\cite{peter;gh21}. This program was used to fit mathematical functions to the asymmetry spectra via non-linear least squares regression, as well as to integrate the asymmetry spectra collected at each temperature and field condition over the full time window. 
	
	\section{Results and Analysis}
	The new results and analysis presented in the following are interspersed with a selection of results published previously~\cite{elmsl;prb22, elmsl;prb23} to facilitate comparison of different samples and highlight emerging trends. New in this work are magnetometry data on the premag sample; ZF $\mu$SR data on the cold-worked and premag samples; LF $\mu$SR data on the pristine sample; analysis of the early-time and time-integrated ZF asymmetry for all samples; and extraction of the spin fluctuation rate and spin-flip activation energy from detailed analysis of the LF data on the pristine, Mn$_0$, and Mn$_{30}$ samples. Results carried over from previous publications include magnetometry data for all but the premag sample; ZF $\mu$SR data and initial anlaysis for the pristine, Mn$_0$, and Mn$_{30}$ samples; and LF $\mu$SR relaxation rates for Mn$_0$ and Mn$_{30}$. The present work builds upon the previously published $\mu$SR results~\cite{elmsl;prb22, elmsl;prb23} by focusing on the microscopic details of the magnetic transition and dynamics, including homogeneity, spin fluctuation rates and energies, and the tunability of these properties via composition, sample processing, and magnetic field. 
	
	\subsection{SQUID Magnetometry}
\begin{figure}
		\includegraphics[width=70mm]{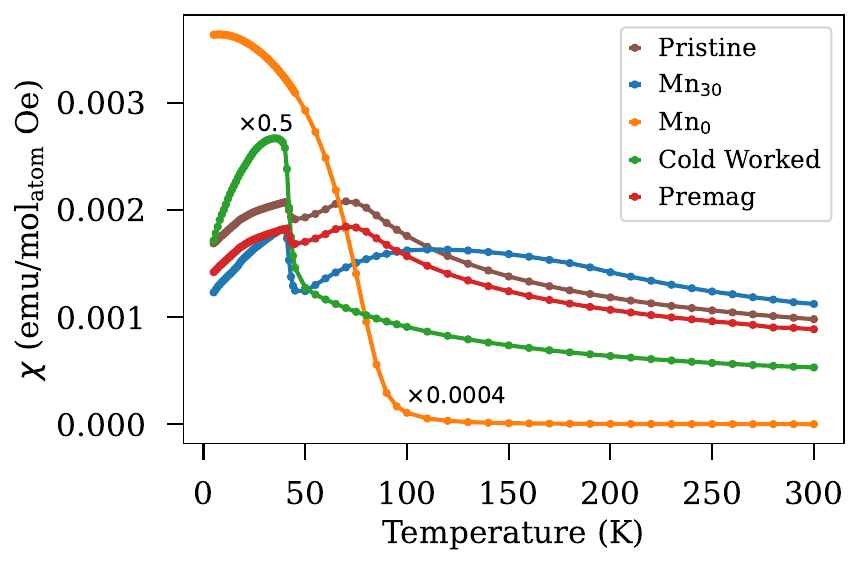}
		\caption{\label{fig:mag}
			Temperature-dependent molar susceptibility for the five Cantor alloys studied in this work. Measurements were performed in a zero-field-cooled warming sequence in an applied field of 100~Oe. Note that the cold-worked and Mn$_{0}$ data have been scaled down by factors of 0.5 and 0.0004, respectively, for convenience. Data for the pristine, Mn$_{30}$, Mn$_0$, and cold-worked samples published previously in~\cite{elmsl;prb22, elmsl;prb23}.}
		\centering
	\end{figure}

	The magnetic susceptibility as a function of temperature as determined by SQUID magnetometry is displayed for each sample in Fig.~\ref{fig:mag}.
	Measurements were conducted in a warming sequence after cooling to 5~K in zero field. The applied field during the measurement was 100~Oe. Detailed discussion of the magnetometry data can be found in Refs.~\onlinecite{elmsl;prb22} and \onlinecite{elmsl;prb23}; here, we briefly note just the most salient features. The pristine, premag, cold-worked, and Mn$_{30}$ samples all show a step-like change around 43~K, attributable to the low-temperature ferrimagnetic transition. The pristine and premag samples additionally show a peak around 70 -- 80~K, which has been established as a signature of the spin-glass-like transition. The cold-worked sample shows no clear peak around that temperature, indicating that plastic deformation tends to suppress the higher transition~\cite{elmsl;prb22}. The Mn$_{30}$ sample also has a peak at high temperature, but it is significantly broader compared to the pristine and premag samples and begins at a much higher temperature (around 180~K), confirming the increase in the spin-glass-like transition temperature with excess Mn~\cite{elmsl;prb23}. Finally, the Mn$_{0}$ sample shows a large increase in magnetic susceptibility with decreasing temperature below about 100~K, indicative of the ferromagnetic ground state demonstrated for Mn-deficient samples~\cite{elmsl;prb23}.
	
	\subsection{Zero-Field \muSR\ Data}

\begin{figure}
		\includegraphics[width=8cm]{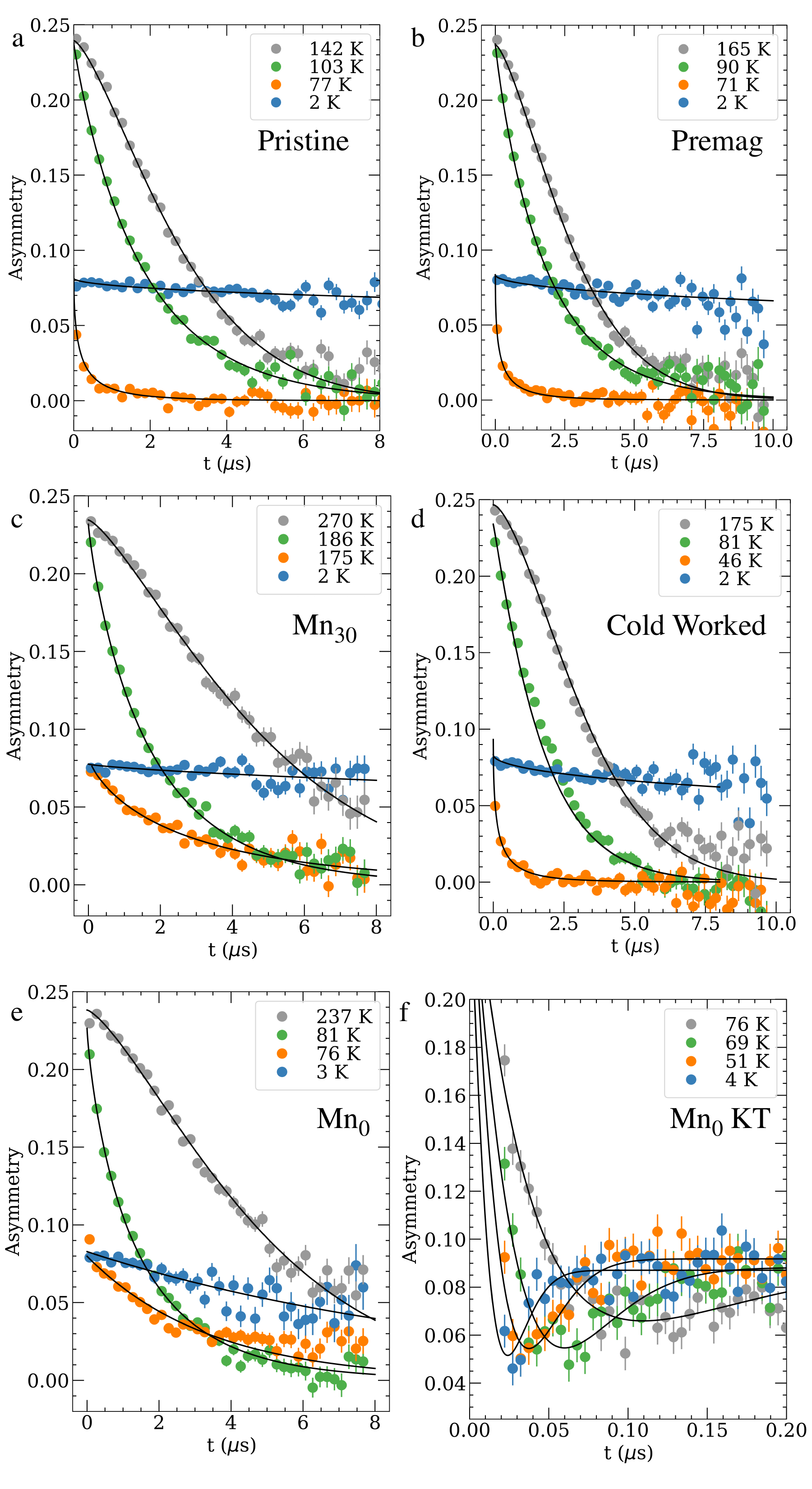}
		\caption{\label{fig:ZFspectra}
			Representative spectra for each Cantor alloy sample (labeled on the individual plots). Filled symbols show the experimental data, and the solid black curves are fits using a stretched exponential function (a-e) or a Kubo-Toyabe function (f). Note the shorter time window visible in panel (f), in contrast to the longer time window used in the other panels. Panels (a), (c), and (e) contain data published previously~\cite{elmsl;prb22, elmsl;prb23}.}
		\centering
	\end{figure}

	We now present the ZF \muSR\ data for the five Cantor alloy samples studied. Representative asymmetry spectra are shown in Fig.~\ref{fig:ZFspectra}.
	For all samples, the relaxation has predominantly Gaussian-like curvature at high temperature (e.g. 142~K for the pristine sample), attributable to the effect of static nuclear dipolar fields. As the temperature decreases, the asymmetry spectra adopt a more exponential lineshape (e.g. 103~K for the pristine sample), indicating that fluctuating electronic moments become the dominant relaxation mechanism. Once the sample is cooled below the spin-glass transition temperature, rapid depolarization of the muon spin ensemble causes the initial asymmetry at $t=0$ to drop to \(1/3\) of the initial asymmetry level at high temperature. This ``\(1/3\) tail'' occurs when a material consists of regions of static magnetization that occupy the full sample volume and are oriented randomly relative to each other~\cite{lee1999muon}, confirming that all the Cantor alloy variants studied here transition to a static magnetic state that extends throughout the entire sample volume at low temperature. Just below the transition (e.g. 77~K for the pristine sample), the 1/3 tail relaxes rapidly due to thermal excitations of the static magnetic state. As the temperature decreases further (e.g. 2~K for the pristine sample), less relaxation is seen in the 1/3 tail, consistent with magnetic fluctuations freezing out at low temperature.
	
	We note that for all samples except Mn$_{0}$, no oscillations are observable in the ``\(2/3\) component'' of the asymmetry at low temperature, which indicates that the local magnetic field probed by the muons is very large and/or has a broad distribution of field strengths~\cite{lee1999muon}. In the case of Mn\(_{0}\), the early-time portion of the spectra below the transition temperature shows a Kubo-Toyabe-type relaxation pattern~\cite{lee1999muon}, as seen in Fig.~\ref{fig:ZFspectra}(f). This behavior points to a narrower distribution of field strengths at the muon stopping sites in Mn$_{0}$ compared to the other samples, which can be rationalized by the reduced number of distinct magnetic species in the Mn$_{0}$ sample. 
	
\begin{figure}
		\includegraphics[width=8cm]{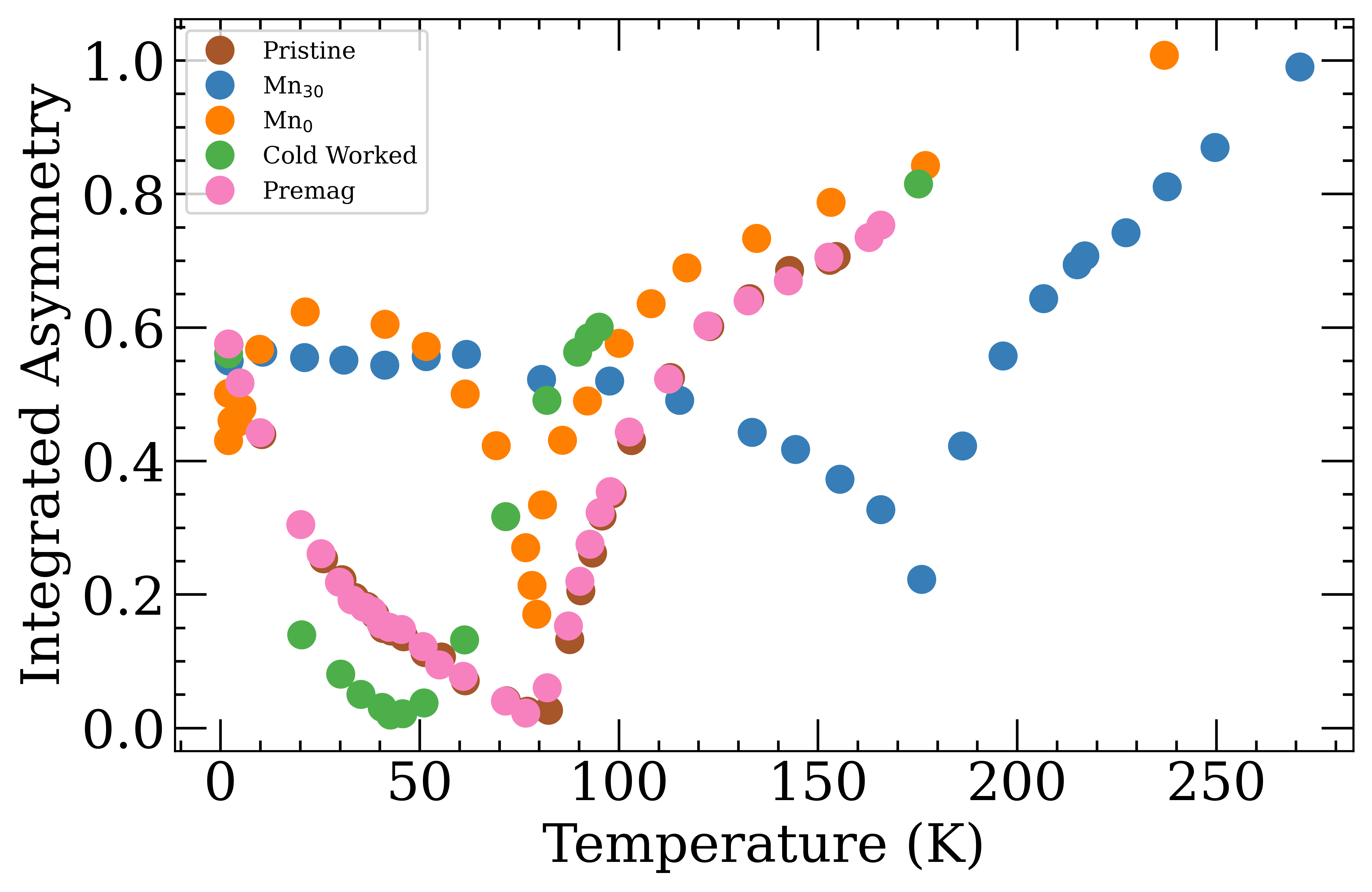}
		\caption{\label{fig:intAsy}Time-integrated ZF asymmetry as a function of temperature for the five Cantor alloy samples studied.}
		\centering
	\end{figure}

	The time-integrated asymmetry can provide a useful average metric for efficient comparison of many different spectra. We display the integrated ZF asymmetry for all five samples in Fig.~\ref{fig:intAsy}.
	The temperature at which the integrated asymmetry is minimized marks the spin-glass transition for each sample, since this is where the 2/3 component vanishes and the 1/3 tail relaxes most rapidly. Based on the results presented in Fig.~\ref{fig:intAsy}, the pristine and premag samples have virtually indistinguishable transitions between 82~K and 87~K, while the cold-worked sample has a lower transition at \(\sim\)50~K. Regarding the non-equiatomic compositions, Mn\(_{30}\) has a higher transition temperature of \(\sim\)175~K, while Mn\(_0\) has a slightly lower transition temperature of \(\sim\)78~K. The cold-worked sample also has broader minimum than do the other samples, which is due to a less homogeneous transition in the cold-worked sample, as will be demonstrated subsequently. The transition temperatures identified from the integrated ZF asymmetry correspond well to the magnetization data and the results published in Refs.~\onlinecite{elmsl;prb22} and \onlinecite{elmsl;prb23}.
	
\begin{figure}
		\begin{center}
			\includegraphics[width=8cm]{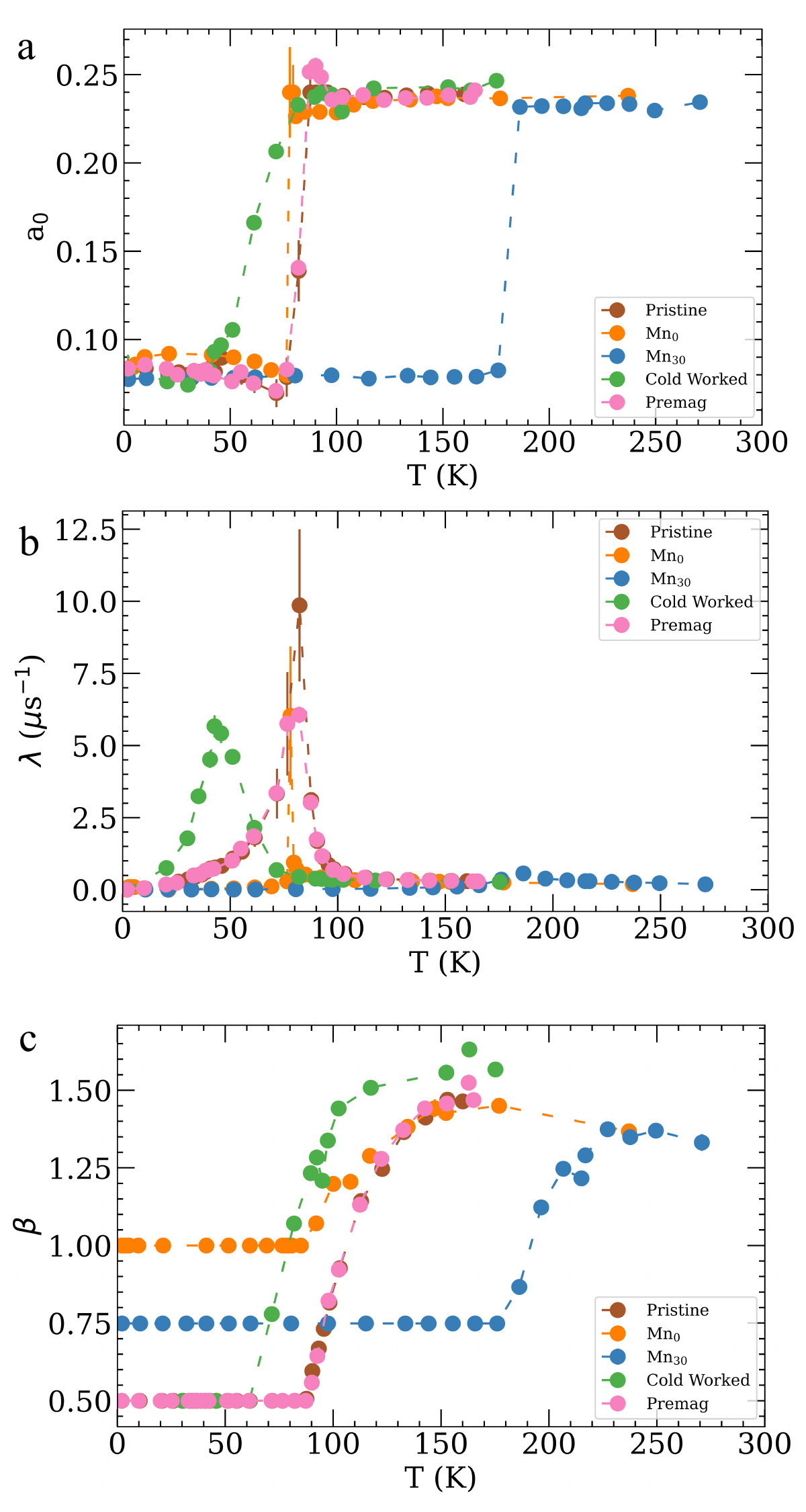}
		\end{center}
		\caption{\label{fig:ZFfitParams}Temperature dependence of the best-fit parameters extracted from fits to the ZF data using Eq.~\ref{eq:strexp}, including (a) the initial asymmetry $a_0$, (b) the relaxation rate $\lambda$, and (c) the exponential power $\beta$. For comparison to the results in (a), the full instrumental asymmetry is approximately 0.24, demonstrating the transition from a completely paramagnetic state at high temperature to static magnetic state occupying the full sample volume at low temperature. Results for the pristine sample were originally reported in Ref.~\onlinecite{elmsl;prb22}, and the results for Mn$_0$ and Mn$_{30}$ in Ref.~\onlinecite{elmsl;prb23}}.
	\end{figure}

	More quantitative information can be obtained by performing least-squares fits to the asymmetry spectra. The solid curves in Fig.~\ref{fig:ZFspectra} show least-squares fits using a stretched exponential function,
	\begin{equation}
		\label{eq:strexp}
		a(t)=a_0e^{-(\lambda t)^\beta},
	\end{equation}
	where \(a(t)\) is the time-dependent asymmetry, \(a_0\) is the initial asymmetry at \(t = 0\), \(\lambda\) is the relaxation rate, and \(\beta\) is the exponential power. This is a phenomenological model typically used in situations where the spectral lineshape is intermediate between purely Gaussian and purely exponential, often associated with a continuous distribution of relaxation rates \cite{crook1997voigtian}. The best-fit parameter values for each sample are shown in Fig.~\ref{fig:ZFfitParams}.
	As seen in Fig.~\ref{fig:ZFfitParams}(a), the initial asymmetry $a_0$ shows a step-like change at the transition temperature for every sample except the cold worked, indicating an abrupt magnetic transition that occurs uniformly throughout the full sample volume within an extremely narrow temperature range. The cold worked sample exhibits the same overall change in the value of $a_0$, but the change occurs over the temperature interval 41~K to 82~K. This demonstrates that different regions of the sample transition at different temperatures in this temperature range and explains why no clear peak was observed in the magnetometry data for this sample.
	
	The temperature dependence of the relaxation rate \(\lambda\) is shown in Fig.~\ref{fig:ZFfitParams}(b). A prominent peak centered around 82~K appears for the pristine and premag samples and 78~K for Mn\(_{0}\). The temperatures around which the peaks are centered correspond well to the transition temperatures already identified for each sample. In the fast fluctuation limit, $\lambda$ is inversely proportional to the spin fluctuation rate~\cite{uemur;prb85}, so an increase in $\lambda$ denotes a decrease in fluctuation rate. Thus, the peak observed in the relaxation rate is evidence of ``critical slowing down'' of the spin dynamics in the vicinity of the transition, as seen in canonical spin glasses and continuous phase transitions \cite{yaouanc2011muon,uemur;prb85}. The cold-worked and Mn\(_{30}\) samples also exhibit a peak in $\lambda$ centered around 43~K and 186~K, respectively, but the peak is much broader for the cold-worked sample and much smaller in magnitude for Mn\(_{30}\). The broadness in the cold-worked sample is attributable to the wide spread of transition temperatures throughout the sample volume, while the smaller peak observed for Mn\(_{30}\) reveals a suppression of the critical spin dynamics in this sample compared to the others.
	
	The \(\beta\) values extracted from the ZF fits are shown in Fig.~\ref{fig:ZFfitParams}(c). For all samples, the values at high temperature are significantly above unity, consistent with the somewhat Gaussian-like curvature observed in Fig.~\ref{fig:ZFspectra}. With decreasing temperature, $\beta$ decreases until it becomes approximately constant below the transition temperature. In our initial fits, we allowed $\beta$ to vary independently for each run, and we observed that the values naturally converged to approximately 0.5 for the pristine, premag, and cold-worked samples, 0.75 for Mn$_{30}$, and 1.0 for Mn$_0$, with some scatter in the results on the order of 10\%. For simplicity, we then conducted another set of fits with $\beta$ fixed to these values below the transition. The value of $\beta$ is physically significant: in situations with a single relaxation channel with just one relaxation rate, $\beta$ is expected to be unity, while the effect of multiple relaxation channels and/or a distribution of relaxation rates is typically modeled with $\beta < 1$~\cite{uemur;prb85, rover;prb02, dunsi;prb96}. The sub-exponential behavior observed in the equiatomic and Mn$_{30}$ samples is similar to the behavior seen in a five-component high-entropy oxide~\cite{frand;prm20}, suggesting this property may be a general feature of \muSR\ spectra in magnetic high-entropy materials. The configurational entropy is maximized for the equiatomic composition and decreases as the composition moves further away from equal proportions. As such, the systematic increase of the low-temperature $\beta$ value from 0.5 in the equiatomic samples to 0.75 in Mn$_{30}$ to 1.0 in Mn$_{0}$ further supports the notion that increased configurational entropy drives the sub-exponential relaxation.
	
	\subsection{Longitudinal-Field \muSR\ Data}

\begin{figure}
		\includegraphics[width=8cm]{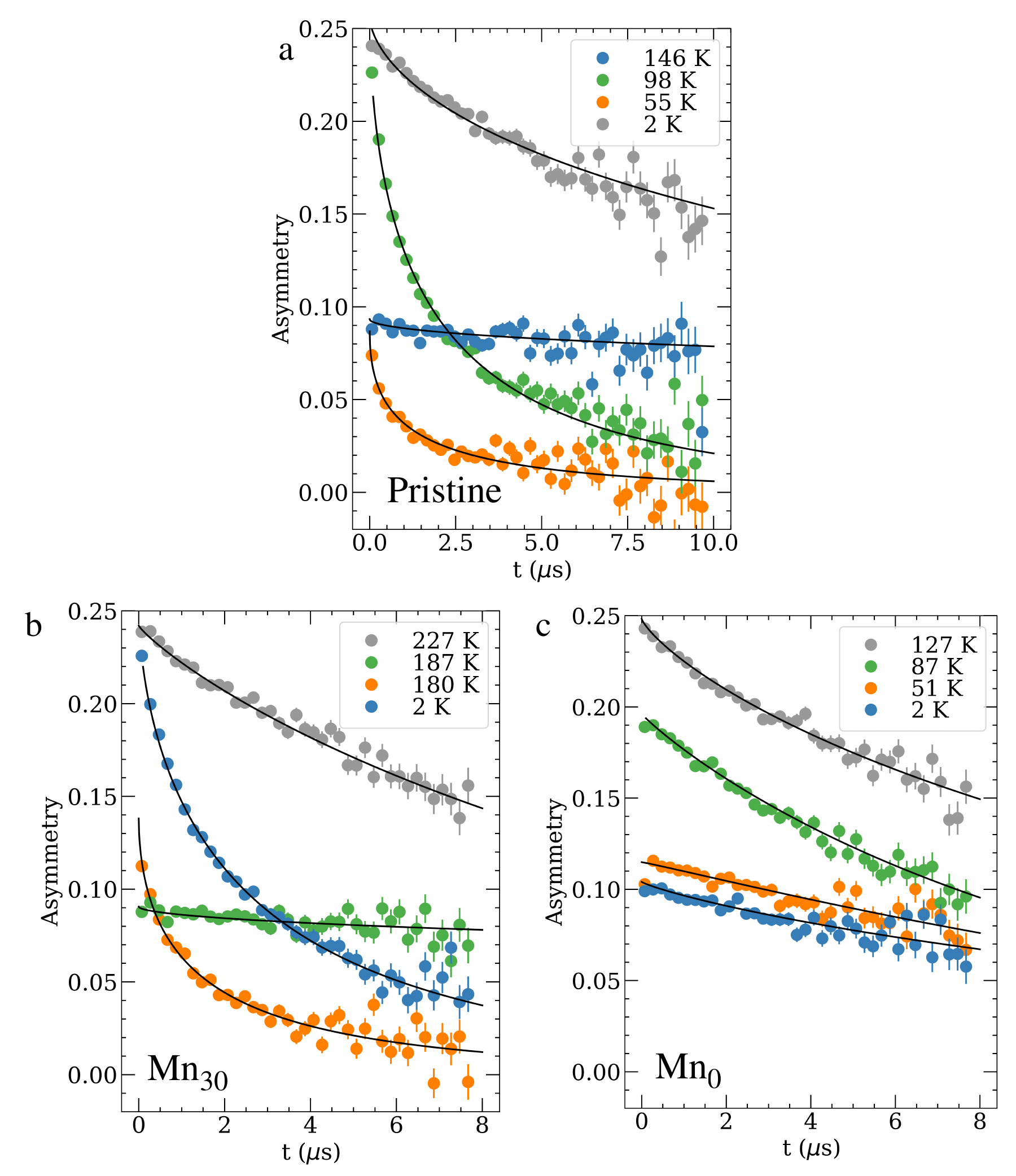}
		\caption{\label{fig:LFspectra}Representative LF \muSR\ spectra for the pristine, Mn$_{0}$, and Mn$_{30}$ samples for an applied 0.1~T field. Filled symbols represent the experimental data, the black curves the fits using a stretched exponential function.}
		\centering
	\end{figure}

	To probe the spin dynamics in more detail, we performed LF \muSR\ experiments, in which the applied field is directed parallel to the initial muon spin polarization. If the internal field at the muon site is static and weak compared to the externally applied field, as is the case for nuclear dipolar fields, then the total magnetic field (i.e., the vector sum of the applied and internal fields) is dominated by the applied longitudinal field, such that minimal muon spin precession occurs and the \muSR\ asymmetry is largely constant. On the other hand, if the internal field is fluctuating and comparable in magnitude to the applied field, then the \muSR\ asymmetry decays with time due to muon spin precession in the rapidly changing local field. Thus, the observed relaxation can be related directly to the underlying spin dynamics without complications from nuclear dipolar fields that arise in ZF data. In this study, we applied a longitudinal field of 0.1~T and measured the asymmetry spectra at several temperatures across the magnetic transition for the pristine, Mn$_{0}$, and Mn$_{30}$ samples. Representative spectra are shown in Fig.~\ref{fig:LFspectra}.
	As with the ZF data, we observe slow relaxation at high temperature (although without the Gaussian curvature stemming from the nuclear dipolar fields) with increasing relaxation as the temperature decreases toward the transition. The initial asymmetry drops substantially below the transition, and the relaxation of the long-time tail is most pronounced near the transition. Qualitatively, these results confirm the presence of static magnetism at low temperature and robust spin fluctuations in a wide temperature range spanning the transition temperature.
	
\begin{figure}
		\includegraphics[width=8cm]{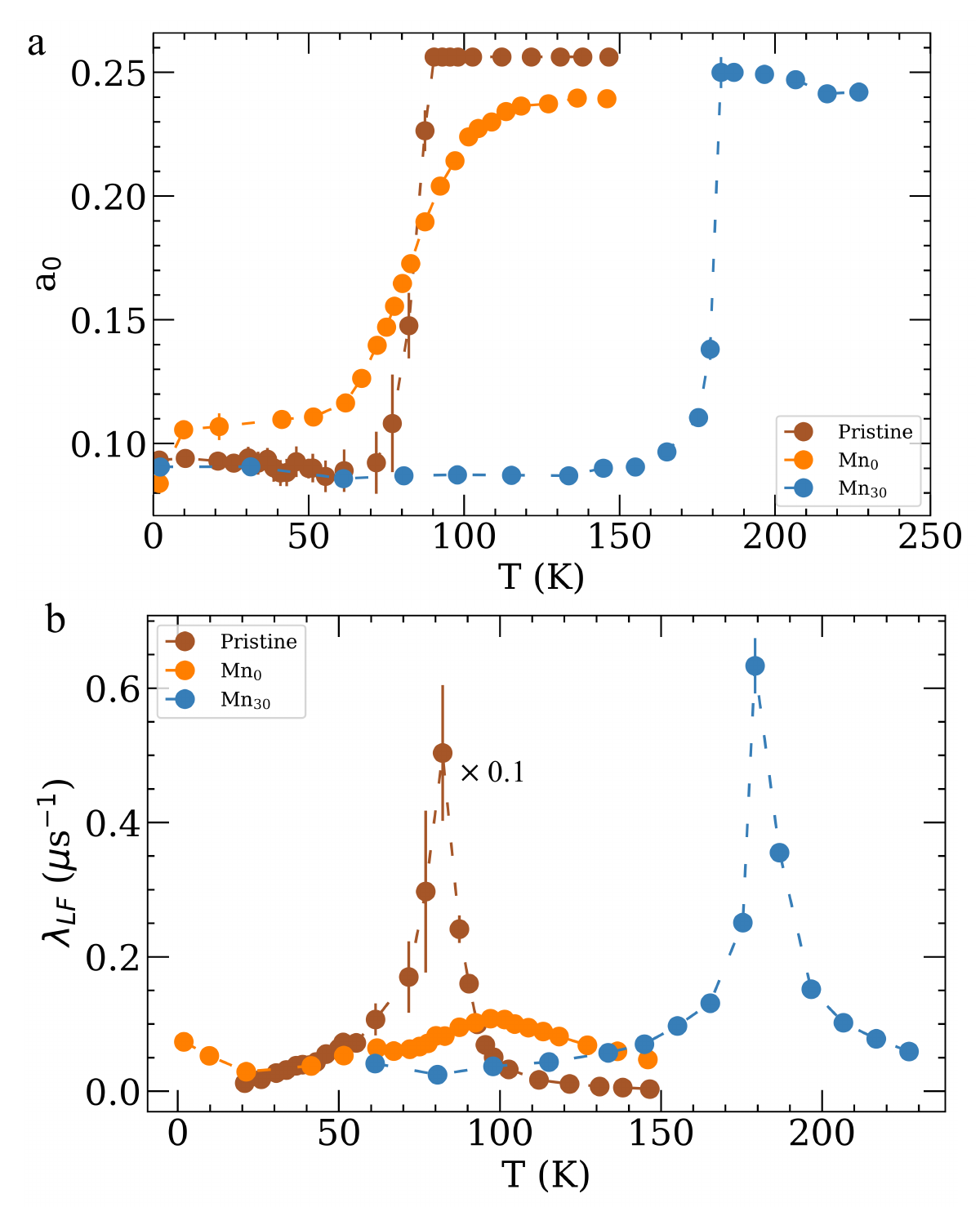}
		\caption{\label{fig:LFparams}Temperature dependence of the best-fit parameters extracted from fits to the LF data using Eq.~\ref{eq:strexp}, including (a) the initial asymmetry $a_0$ and (b) the relaxation rate $\lambda_{LF}$. Note that $\lambda_{LF}$ for the pristine sample has been divided by 10 for ease of viewing. $\lambda_{LF}$ results for the Mn$_0$ and Mn$_{30}$ samples were first published in~\cite{elmsl;prb23}.}
		\centering
	\end{figure}

	We modeled the data once again with a stretched exponential as in Eq.~\ref{eq:strexp}, where we use $\lambda_{LF}$ to distinguish the relaxation rate from the ZF case. Fig.~\ref{fig:LFparams}(a) shows the initial asymmetry $a_0$ obtained from the fits.
	A rapid change is observed at the transition temperature for the pristine and Mn$_{30}$ samples, confirming the presence of a uniform magnetic transition throughout the full sample volume. In contrast, the Mn\(_{0}\) sample shows a much more gradual change between about 50~K and 120~K, indicating that the application of a magnetic field greatly broadens the transition in this sample. This result is consistent with the predominantly ferromagnetic behavior expected in this sample, wherein the enhanced magnetic susceptibility allows the external field to magnetize local regions of the sample at temperatures well above the transition in zero field. With the more balanced antiferro- and ferromagnetic interactions in the pristine and Mn$_{30}$ samples, the external field has a smaller effect on the transition temperature.
	
	The long-time relaxation rate \(\lambda_{LF}\) is displayed in Fig.~\ref{fig:LFparams}(b). The pristine sample features a large, sharp peak centered around 82~K, confirming that a critical slowing down of the spin fluctuations occurs in the vicinity of the transition temperature. Mn\(_{30}\) likewise shows a sharp peak centered around 183~K, although the magnitude is much smaller than that of the pristine sample. In contrast, Mn\(_{0}\) shows a broad, low peak centered around 97~K. The breadth of the peak is due to the smeared out transition in an applied field, while the diminished height of the LF peak indicates that the critical slowing down observed in the pristine sample and (to a slightly lesser extent) the Mn$_{30}$ sample is greatly suppressed in the Mn$_{0}$ sample. These aspects of the peak shape differ from the ZF data for Mn$_{0}$, which exhibits a sharp peak in the relaxation rate at the transition temperature, suggesting the application of a relatively modest field of 0.1~T significantly modifies the internal spin dynamics. We note that the shift of the peak center from 78~K in ZF to 97~K in LF is due to ferromagnetic nature of Mn$_0$, in contrast to the other compositions, which do not show a large change in the temperature of the peak center with applied field.
	
\begin{figure}[t]
		\begin{center}
			\includegraphics[width=8cm]{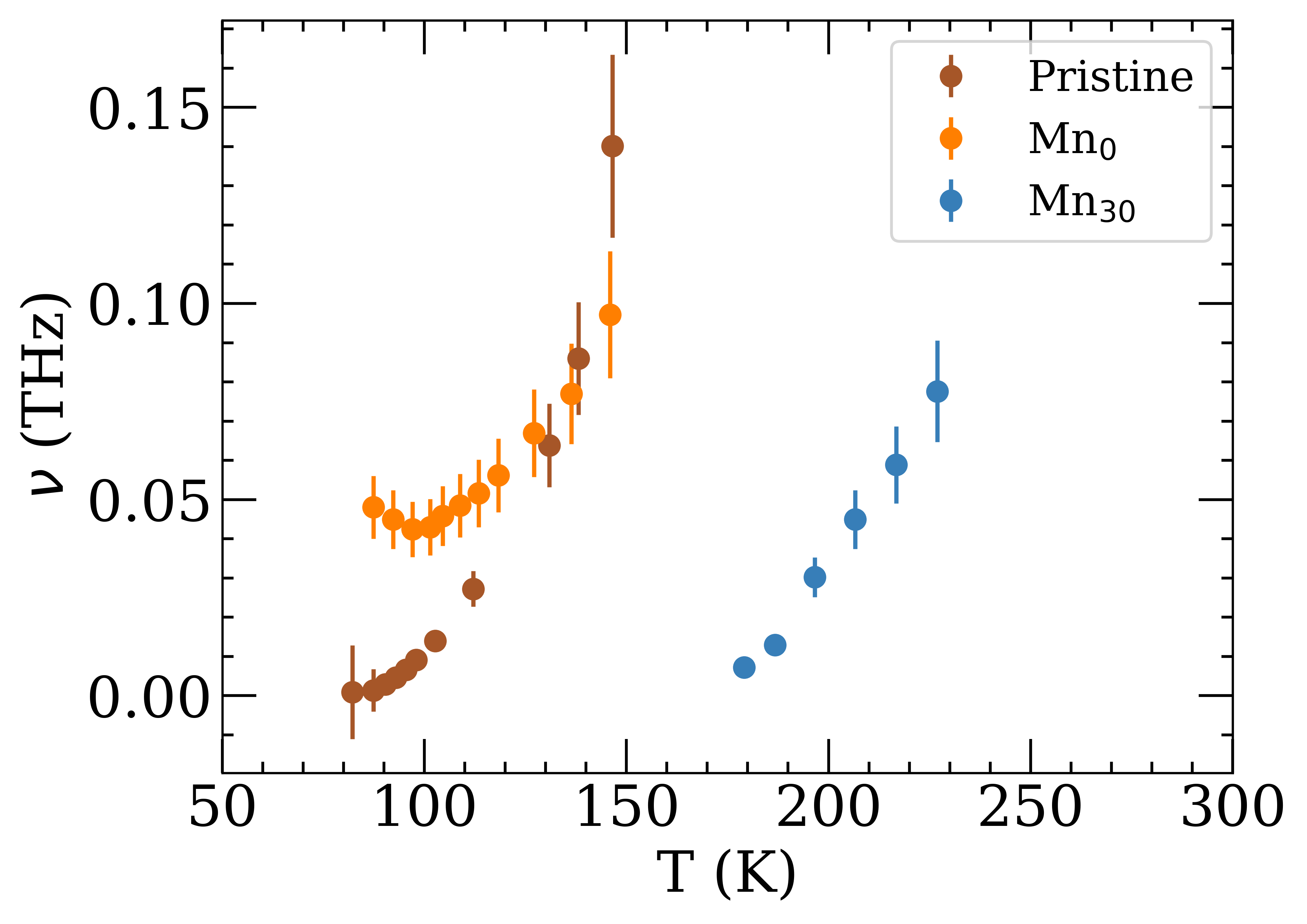}
		\end{center}
		\caption{\label{fig:nu} Spin fluctuation rate for the pristine, Mn$_{0}$, and Mn$_{30}$ samples extracted from the LF \muSR\ fits. The decrease approaching the respective transition temperature for each sample is characteristic of the critical slowing down of the spin fluctuations near the transition.}
	\end{figure}

	The difference in critical spin dynamics between the three samples can also be demonstrated by estimating the spin fluctuation rate $\nu$ from the LF data.
	In a system with a single relaxation channel in the paramagnetic phase, the relaxation rate \(\lambda_{LF}\) is related to \(\nu\) through~\cite{uemur;prb85}
	\begin{equation}
		\label{eq:nu}
		\nu = \frac{\Delta^2 +  \sqrt{\Delta^4 -\lambda_{LF}^2\omega_L^2}}{\lambda_{LF}},
	\end{equation} where \(\Delta = \gamma_\mu B_i\) is the product of the muon's gyromagnetic ratio $\gamma_{\mu}$ and the root mean square internal field \(B_i\), and \(\omega_{L} = \gamma_\mu B_L\) is the Larmor frequency of the muon spin in the longitudinal field \(B_L\). As has been done elsewhere~\cite{frand;prm20}, \(\Delta\) can be estimated from the Kubo-Toyabe relaxation rate observed at low temperature, which is $120 \pm 10$~$\mu$s$^{-1}$ for Mn\(_{0}\). Since the pristine and Mn$_{30}$ samples do not exhibit a clear Kubo-Toyabe relaxation pattern, we use the value of $\Delta$ determined from the Mn$_0$ data as an upper bound when calculating $\nu$. We base this on the fact that the net effective moment increases as the Mn content is decreased~\cite{elmsl;prb23}, leading to a larger expected local field for Mn$_0$. Thus, the calculated values of $\nu$ for the pristine and Mn$_{30}$ samples should be considered upper bounds. We further note that Eq.~\ref{eq:nu} relies on the assumption that the asymmetry exhibits purely exponential relaxation in the paramagnetic phase, which is only approximately the case in the present study, but this calculation nevertheless yields a reasonable estimate of the spin fluctuation rate as a function of temperature. The calculated spin fluctuation rate is shown for the pristine, Mn\(_0\) and Mn\(_{30}\) samples in Fig.~\ref{fig:nu}. 
	For the pristine sample, $\nu$ approaches zero as $T$ decreases toward the transition temperature, directly confirming the critical slowing down of spin fluctuations in the vicinity of the transition. Mn$_{30}$ likewise shows a rapid drop in $\nu$ with decreasing temperature, although the minimum spin fluctuation rate before the transition is higher than that in the pristine sample. For Mn$_{0}$, $\nu$ stays relatively constant around 0.05~THz below 100~K, confirming that the critical slowing down is partially suppressed in this sample. We note that the fluctuation rates observed here are similar in magnitude to those observed in a high-entropy oxide~\cite{frand;prm20} and canonical spin glasses~\cite{uemur;prb85}.
	
\begin{figure}
		\begin{center}
			\includegraphics[width=8cm]{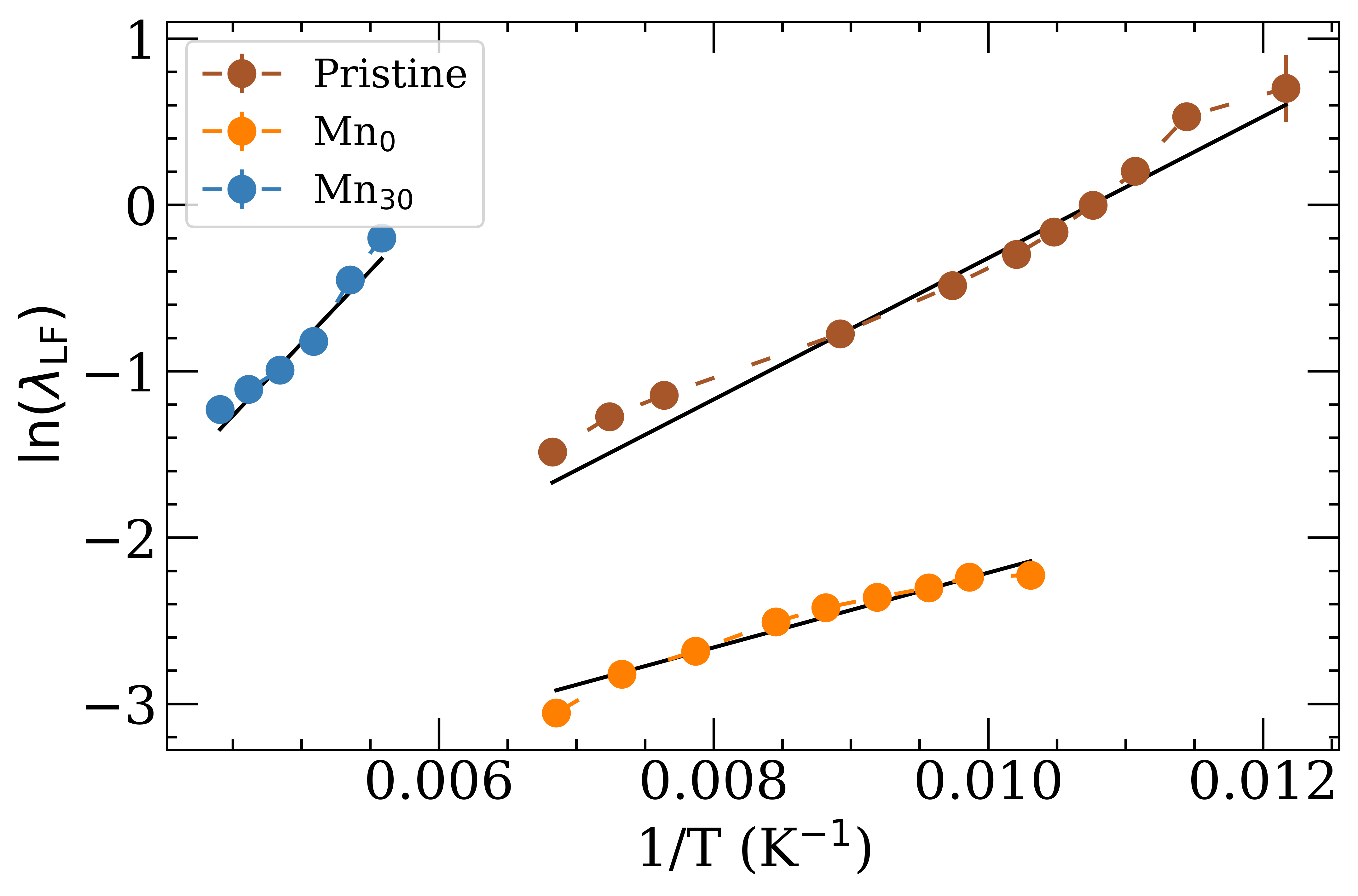}
		\end{center}
		\caption{\label{fig:loglambda}Logarithm of the LF relaxation rate versus inverse temperature for the pristine, Mn$_0$, and Mn$_{30}$ samples. The solid lines are linear fits to the data, with the slope representing the spin-flip activation energy, as described in the main text.}
	\end{figure}

	
	Further insight into the spin dynamics can be gleaned by analyzing the temperature dependence of $\lambda_{LF}$. The behavior above the transition temperature can be well described by a thermal activation model that has been used successfully for \muSR\ studies of magnetic nanoparticles~\cite{rebbo;prb07, frand;prm21}. In this model, the temperature-dependent relaxation rate is given by
	\begin{equation}
		\label{eq:lambdaLF}
		\lambda_{LF}(T) = \lambda_0 \exp\left(E_a/k_BT\right),
	\end{equation} where $T$ is the temperature, \(E_a\) is the activation energy for flipping a spin, \(k_B\) is the Boltzmann constant, and \(\lambda_0\) is an intrinsic relaxation rate. In Fig.~\ref{fig:loglambda}, we plot the logarithm of \(\lambda_{LF}\) versus inverse temperature, which results in a linear plot for an ideal thermal activation scenario.
	The experimental results indeed follow a linear trend, confirming the appropriateness of this model. We performed a linear fit to the data for each sample (solid lines in Fig.~\ref{fig:loglambda}) to extract the slope, which corresponds to \(E_A/k_B\). From this analysis, we determined the slopes to be $430 \pm 30$~K, $220 \pm 20$~K, and $870 \pm 120$~K for the pristine, Mn$_0$, and Mn$_{30}$ samples, respectively, corresponding to activation energies of $5.9(4) \times 10^{-21}$~J, $3.1(3) \times 10^{-21}$~J, and $1.2(2) \times 10^{-20}$~J, respectively. We note that the slope increases monotonically with Mn content, suggesting that the average spin-flip activation energy increases with increasing antiferromagnetic character associated with the Mn atoms. In addition, the activation energies are within a factor of $\sim$5 of the transition temperature for each sample, a similar ratio to that observed in iron-oxide magnetic nanoparticles~\cite{rebbo;prb07,frand;prm21}.
	
	\section{Discussion and Conclusion}
	We have presented comprehensive ZF and LF \muSR\ studies of several Cantor alloy variants, including three equiatomic samples with different sample processing histories and two non-equiatomic samples with varied Mn content. The results shed light on the magnetic transitions and underlying spin dynamics in these samples, demonstrating that the magnetic properties can be substantially modified through the choice of composition and sample processing. Excess Mn significantly increases the spin-glass transition temperature relative to the equiatomic compound, while excluding Mn altogether results in a ferromagnetic transition at slightly lower temperature, confirming previous results~\cite{elmsl;prb23}. Cold working the equiatomic compound dramatically alters the transition, resulting in an inhomogeneous transition that develops gradually throughout the sample volume over a wide temperature range, in contrast to the sharp transition that occurs uniformly throughout the sample volume in non-cold-worked samples. This difference in the cold-worked sample suggests that plastic deformation modifies the atomic and magnetic homogeneity on the micro- or nanoscale to the point that different regions of the sample transition at different temperatures. More detailed studies using scanning electron probes could provide greater clarity on this issue. On the other hand, magnetizing the equiatomic Cantor alloy in a field of 9~T had no observable effect on the magnetic properties.
	
	Perhaps the most interesting findings from the present study relate to the spin dynamics in Cantor alloys. We observed critical spin dynamics near the magnetic transition in every sample studied, manifest by a peak in the \muSR\ relaxation rate at the transition temperature. This type of behavior is commonly observed in canonical spin glasses and systems with ideal continuous phase transitions~\cite{yaouanc2011muon,uemur;prb85}, so its presence in Cantor alloys despite the extreme chemical disorder is notable. One might naively expect the intrinsic disorder in HEAs to result in distinct nucleation sites of magnetic order and therefore favor a first-order transition, for which critical spin dynamics would not necessarily play a role; however, that is evidently not the case for Cantor alloys, considering the critical behavior observed here. Nevertheless, the details of the critical behavior can be tuned via chemical composition, with larger spin-flip activation energies observed in samples with higher Mn concentration and correspondingly stronger antiferromagnetic character. In addition, an applied magnetic field suppresses the critical spin dynamics for the Mn$_0$ sample, though not for the equiatomic and Mn$_{30}$ samples. We likewise attribute this to the predominantly ferromagnetic character of the Mn$_0$ sample, with its large magnetic susceptibility, in contrast to the Mn$_{30}$ sample with enhanced antiferromagnetic character. The spin fluctuation rates near the transition, i.e. $\sim$0.1~THz, are comparable to those observed via \muSR\ in conventional bulk magnetic materials~\cite{dunsi;prb96,koda;prb04}, and the spin-flip activation energies, i.e. $\sim 10^{-21}$~J, are similar to iron oxide magnetic nanoparticles~\cite{rebbo;prb07,frand;prm21}. Further theoretical and computational work may be necessary to establish the mechanism by which compositional variations influence the details of the spin dynamics.
	
	Overall, this work has provided valuable insights into the spin dynamics of Cantor alloys. The results demonstrate the tunability of the magnetic transitions and dynamics using chemical composition and sample processing as control knobs, which may enable novel types of magnetic engineering using Cantor alloys in the future. The findings also invite further studies of other magnetic HEAs using \muSR\ and similar local magnetic probes to help establish which behaviors are generic to HEAs and which are specific to individual systems.
	
	\textbf{Acknowledgements}
	
	We thank the staff at TRIUMF for their valuable help and support during the \muSR\ experiments, along with Raju Baral, Braedon Jones, Kane Fanning, Karolyn Mocellin, Christiana Suggs, and Parker Hamilton for assistance with the \muSR\ data collection. E.Z. and B.A.F. acknowledge the College of Physical and Mathematical Sciences at Brigham Young University for funding support. T.A.E. and M.W.M. acknowledge support from NSF No.
	DMR-1708410. Synthesis and characterization facilities at the University of Florida were developed under support from NSF-CAREER 1453752 (J.J.H.). RD is supported by the Center for Integrated Nanotechnologies, an Office of Science user facility operated for the U.S. Department of Energy. This article has been authored by an employee of National Technology \& Engineering Solutions of Sandia, LLC under Contract No. DE-NA0003525 with the U.S. Department of Energy (DOE). The United States Government retains and the publisher, by accepting the article for publication, acknowledges that the United States Government retains a non-exclusive, paid-up, irrevocable, world-wide license to publish or reproduce the published form of this article or allow others to do so, for United States Government purposes. The DOE will provide public access to these results of federally sponsored research in accordance with the DOE Public Access Plan https://www.energy.gov/downloads/doe-public-access-plan .

\end{document}